\documentclass[twocolumn,superscriptaddress,showpacs,amsmath,amssymb,aps]{revtex4}
\usepackage{graphicx}
\usepackage{amsmath}
\usepackage{multirow}
\usepackage{color}
\usepackage{bm}

\newcommand{\diff}{{\mathrm d}}
\newcommand{\rmc}{{\mathrm c}}

\begin{document}

\title{Monte Carlo simulation with aspect ratio optimization: \\ Anomalous anisotropic scaling in dimerized antiferromagnet}

\author{Shinya Yasuda}
\affiliation{Department of Applied Physics, the University of Tokyo, Tokyo 113-8656, Japan}
\author{Synge Todo}
\affiliation{Institute for Solid State Physics, the University of Tokyo, 7-1-26-R501 Port Island South, Kobe 650-0047, Japan}

\date{\today}

\begin{abstract}
We present a method that optimizes the aspect ratio of a spatially
anisotropic quantum lattice model during the quantum Monte Carlo
simulation, and realizes the virtually isotropic lattice
automatically.  The anisotropy is removed by using the Robbins-Monro
algorithm based on the correlation length in each direction.  The
method allows for comparing directly the value of critical amplitude
among different anisotropic models, and identifying the universality more
precisely.  We apply our method to the staggered dimer
antiferromagnetic Heisenberg model and demonstrate that the apparent
non-universal behavior is attributed mainly to the strong size
correction of the effective aspect ratio due to the existence of the
cubic interaction.
\end{abstract}


\maketitle

\begin{figure}[tbp]
\begin{center}
\includegraphics[bb=17 166 1013 629,width=85mm]{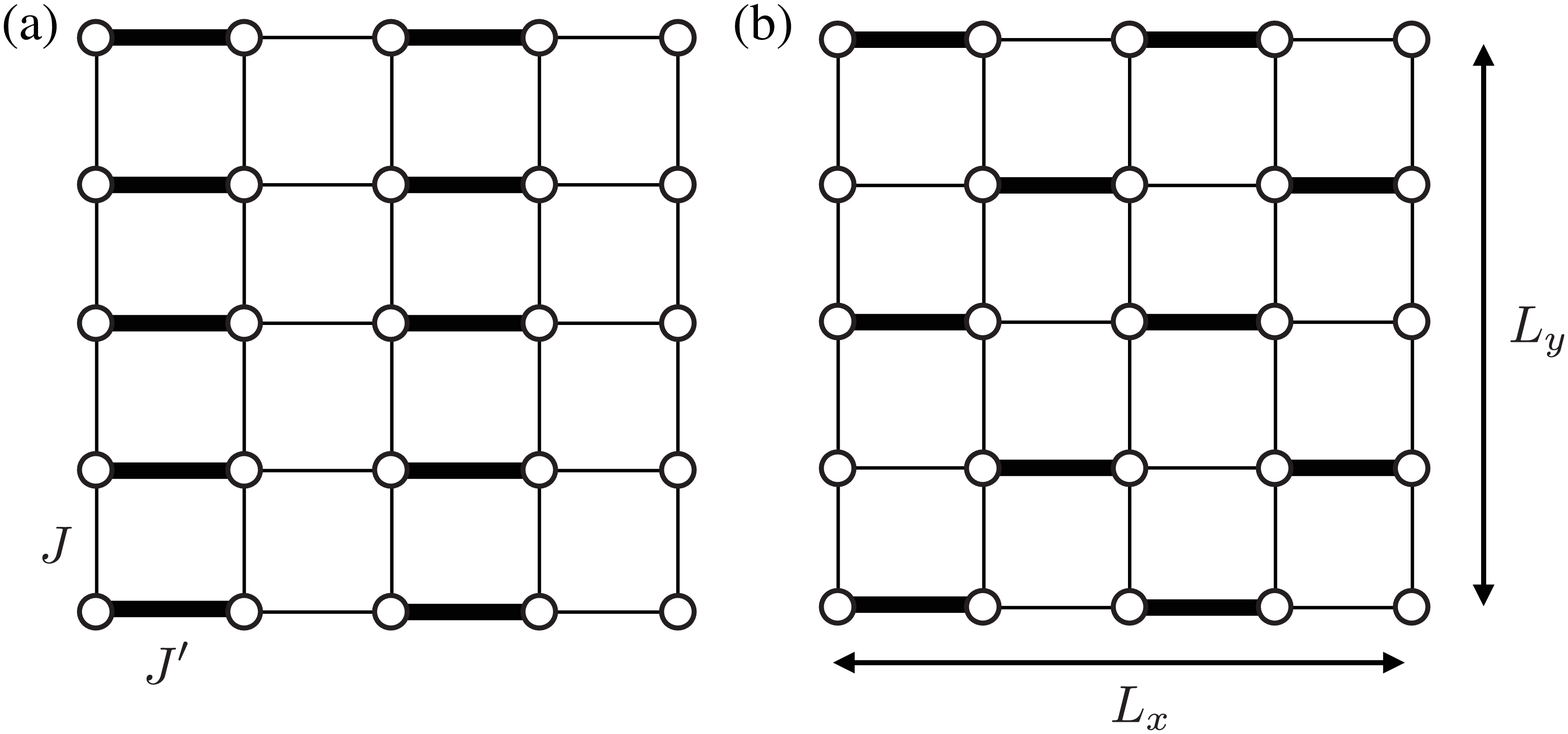}
\caption{Anisotropic antiferromagnetic Heisenberg models: (a) columnar dimer model and (b) staggered dimer model. The thick and thin bonds have the coupling constants, $J^{\prime}$ and $J$, respectively. The linear extent of the system in the horizontal (vertical) direction is denoted as $L_x$ ($L_y$).  Periodic boundary conditions are assumed both in $x$ and $y$-directions.}
\label{models}
\end{center}
\end{figure}

Quantum phase transitions~\cite{Sachdev1999} are the transitions
between ground states with different symmetries.  They are triggered
at absolute zero temperature by the change of a parameter that controls
the strength of quantum fluctuations.  A quantum phase transition in $d$
dimensions, if it is of second order, is widely considered to belong to the
same universality class as the finite-temperature phase transition of the $(d+1)$-dimensional classical system with the same symmetry.
As a concrete example, let us consider
the columnar dimer model, a spin-1/2 dimerized
Heisenberg antiferromagnet.  The Hamiltonian of this system is
written as
\begin{equation}
\mathcal{H} = J \!\! \sum_{\langle i, j\rangle\in A} \!\! \vec{S} _i \cdot \vec{S} _j + J^\prime \!\! \sum_{\langle i, j\rangle\in B} \!\! \vec{S}_{i}\cdot \vec{S}_{j},\label{hamiltonian}
\end{equation}
where $\vec{S}_i$ is the spin-1/2 operator at site $i$, $J$ and
$J^{\prime}$ are the positive (antiferromagnetic) coupling constants, and
$A$ ($B$) denotes the set of the pairs of sites connected by
the thin (thick) bond shown in Fig.\,\ref{models}(a). It undergoes a
second-order quantum phase transition at some critical value of
$J^\prime/J$.  By increasing $J^\prime/J$ from 1, the ground state
changes from the N\'{e}el ordered state to the dimer
state~\cite{MatsumotoYTT2001, WenzelJ2009}, and the critical exponents
of this transition are known to coincide with those of the
three-dimensional (3D) classical Heisenberg [$O(3)$]
universality~\cite{ChakravartyHN1988,Haldane1988,ChubukovSY1994,Vojta2003}.

In the meantime, however, there are a number of intensive researches that aim to find novel critical phenomena that have no classical counterparts.
As for the lattice spin models, the staggered dimer model has been examined as a candidate that might exhibit such phenomena.  Its Hamiltonian is the same as Eq.\,\eqref{hamiltonian}, but
the only difference between the columnar and staggered dimer models is the configuration of the dimerization pattern [Fig.\,\ref{models}(b)].
In Ref.\,\citenum{WenzelBJ2008}, based on the results of the quantum Monte Carlo simulation the authors claim that the critical exponents of the staggered dimer model are different from those of the other dimerized models, such as the columnar dimer model.
In the more recent study~\cite{Jiang2012}, on the other hand, it is pointed out that the simulation results become consistent with the conventional $O(3)$ universality by carefully choosing the aspect ratio of the lattice.  It is further discussed that the models with specific dimerized patterns, including the staggered dimer model, could exhibit apparent unconventional critical phenomena due to the presence of the weakly irrelevant ``cubic term''~\cite{FritzDWWBV2011,KaoTJ2012}, though the relation between such cubic term and the numerically observed large corrections to scaling is yet unclear. 

Usually, quantum Monte Carlo simulations of quantum critical phenomena
are carried out with a cubic geometry, e.g., $L_x : L_y : L_\tau =
1:1:1$ in (2+1) dimensions.  Here, we denote the linear length of the
system in $\alpha$-direction as $L_\alpha$ ($\alpha=x$, $y$, or
$\tau$). The length in $\tau$-direction means the inverse temperature,
$1/T$.  One should be noticed that in the case where the system has
spatially anisotropic interactions, the correlation lengths,
$\xi_\alpha$'s, generally depend on their directions.  In such a case, it is
natural to introduce the {\em virtual aspect ratio},
$R_x^{-1}:R_y^{-1}:R_\tau^{-1}$, by using the effective system linear
length defined as the inverse of relative correlation length,
$R_\alpha \equiv \xi_\alpha / L_\alpha$.

In principle, results of the finite-size scaling analysis do not depend on
the aspect ratio chosen for a series of simulations as long as sufficiently large
lattices are simulated.  In practice, however, one can simulate effectively larger systems with minimal computational cost by tuning the aspect ratio so that the system
becomes virtually isotropic~\cite{MatsumotoYTT2001}, i.e.,
$R_x^{-1}:R_y^{-1}:R_\tau^{-1} \approx 1:1:1$ instead of $L_x : L_y : L_\tau
= 1:1:1$.  By adopting such a geometry, one can examine
the universality of quantum critical phenomena even more closely,
because not only the critical exponents but also the scaling function of quantities with vanishing scaling dimension becomes universal under the virtually cubic geometry.
For example,
let us consider the Binder ratio that is defined as
\begin{equation}
Q=\frac{\langle (m^z)^2\rangle^2}{\langle (m^z)^4 \rangle},\label{eq.q}
\end{equation}
where $m^z$ is the summation of the $z$-component of spins.
The $y$-intercept of the scaling function of such a quantity,
called the {\em critical amplitude}, is a useful index for
identifying the universality class, because such an amplitude can usually be
calculated with higher accuracy than the critical
exponents~\cite{JankeKV1994, Selke2007,NicolaidesB1988,KamieniarzB1993}.

Last of all, for the case where the virtual aspect ratio of the system changes
gradually as increasing the system size, additional care must be
taken, since it might cause the strong corrections to scaling.  As we
will see below, the staggered dimer model is the very case,
and the result reported in Ref.\,\citenum{WenzelBJ2008} is an artifact due to the strong influence
of the non-trivial system size dependence of the virtual aspect ratio.

Tuning the aspect ratio by hand is generally a difficult and complicated task.
In the present paper, we propose an algorithm that optimizes the aspect ratio during the Monte Carlo simulation automatically. This method enables one to make $R_\alpha$ the same value in all directions in order to simulate the virtually isotropic system. It is also possible to search for the quantum critical point $J^\prime=J^\prime_\rmc$ (we set $J=1$ without loss of generality). For example, in a (2+1)-dimensional system, we solve the equation $R_x=R_y=R_\tau=R_0$ for each fixed $L_x$, where $R_0>0$ is an arbitrarily chosen constant.
In this case, we have three parameters, $L_y$, $L_\tau$, and $J^\prime$, to be determined and three equations, which imply that we can determine all parameters.
Note that the reason why $R_0$ is arbitrarily chosen is that in the thermodynamic limit $J^\prime$ smaller (larger) than the critical value $J^\prime_\rmc$ gives the limit $R_\alpha \to \infty$ $(0)$. Thus any positive finite constant $R_0$ leads $J^\prime(L_x) \to J^\prime_\rmc$ as $L_x \to \infty$. In the present simulation, we use $R_0 = 0.5664$, which is an estimate of the critical amplitude of the 3D classical isotropic Heisenberg model at the critical point ($1/T_\rmc=0.693035(37)$ \cite{ChenF1993}) by the Wolff algorithm \cite{Wolff1989}. This is because if the model considered here belongs to the $O(3)$ universality class, this choice of $R_0$ can reduce the corrections to scaling.

In the present simulation, we adopt the loop algorithm based on the continuous-time path integral representation~\cite{EvertzLM1993,Todo2013a}.  The correlation length in each direction is evaluated by the second-moment method~\cite{CooperFP1982,TodoK2001} as
\begin{equation}
  \xi_\alpha = \frac{1}{|\delta\vec{q}_\alpha|}
  \sqrt{\frac{C(\vec{q}_0)}{C(\vec{q}_0 + \delta\vec{q}_\alpha)} - 1},
\end{equation}
where $C(\vec{q})$ is the imaginary-time dynamical structure factor of
the $z$-component of the magnetization at wavevector $\vec{q}$,
$\vec{q}_0 = (\pi,\pi,0)$, and $\delta\vec{q}_\alpha =
(2\pi/L_x,0,0)$, $(0,2\pi/L_y,0)$, and $(0,0,2\pi/L_\tau)$ for $\alpha
= x$, $y$, and $\tau$, respectively.  As the estimates fluctuate
statistically, the naive Newton method becomes unstable and does not
work well for the present purpose.  Instead, we employ a more robust
method, the Robbins-Monro algorithm~\cite{RobbinsM1951,Bishop2006},
from the field of machine learning.  This algorithm enables us to
estimate the zero of the regression function with probability unity
for an observable with a finite variance.

\begin{figure}[t]
\begin{center}
\includegraphics[width=90mm]{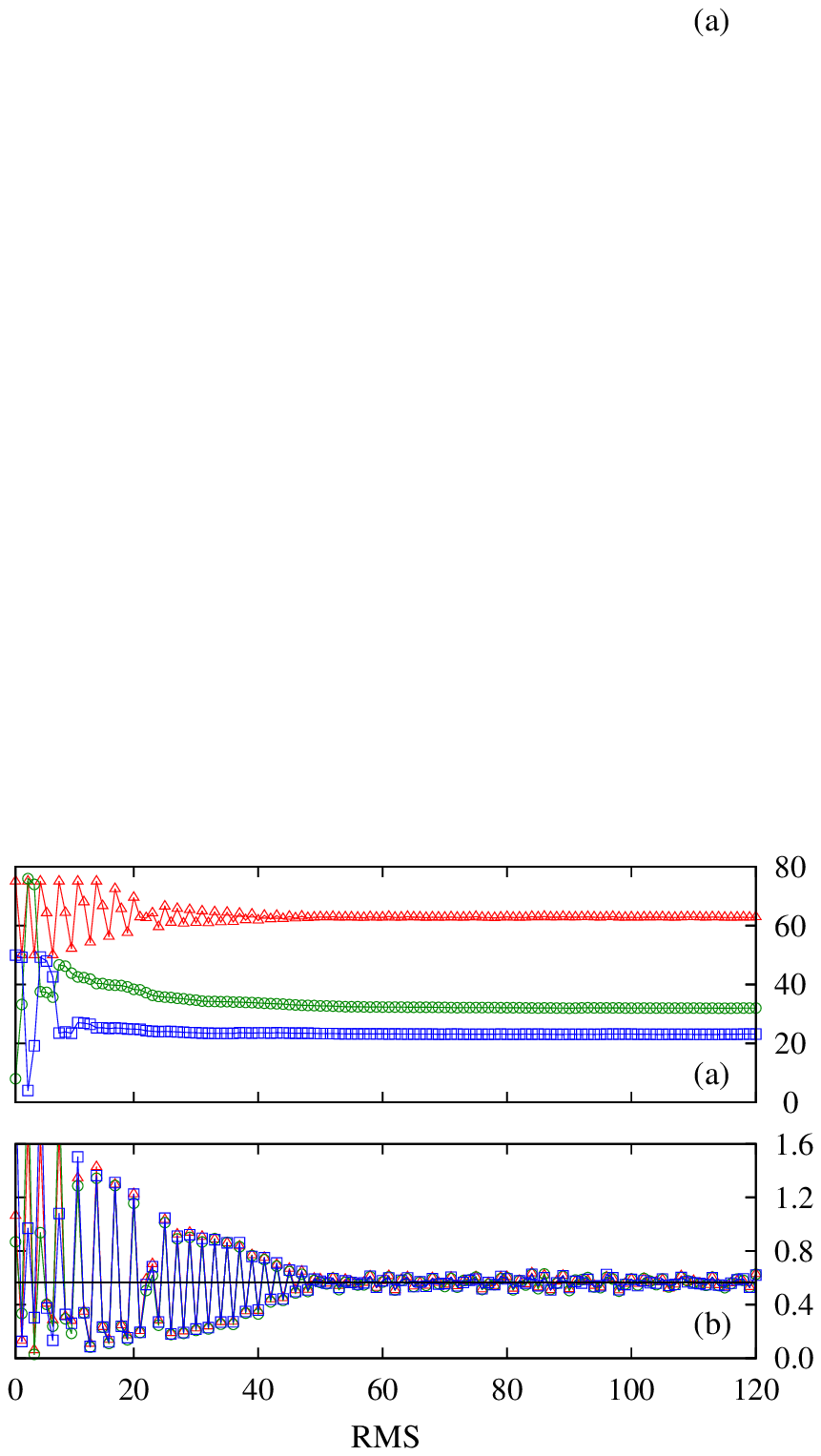}
\caption{Convergence of the parameters and the relative correlation lengths for the staggered dimer model with $L_x=64$: (a) $J'(\times 20)$ (red triangles), $L_y$ (green circles), and $L_\tau$ (blue squares). (b) $R_x$ (red triangles), $R_y$ (green circles), $R_\tau$ (blue squares).  The horizontal axis denotes the Robbins-Monro steps.  The black horizontal line in the lower panel indicates the target value, $R_0 = 0.5664$.}
\label{params_multi}
\end{center}
\end{figure} 

Let $z(\theta)$ be a random variable parameterized by $\theta$ with
mean $f(\theta)$ and a finite variance.  We assume that the regression
function $f(\theta)$ increases monotonically as increasing $\theta$
and has a zero, $\theta = \theta^*$. The zero $\theta^*$ can be
obtained by repeating the Robbins-Monro procedure:
\begin{equation}
\theta^{(n+1)} = \theta^{(n)} - \frac{\alpha}{n} z (\theta^{(n)}),\label{eq4}
\end{equation}
where $n=0,1,2,\cdots$ is the iteration step, $\alpha$ some positive
constant, and $\theta^{(n)}$ the estimate of $\theta^*$ at step $n$.
It is proved that $\theta^{(n)}$ converges to $\theta^*$ with probability one~\cite{RobbinsM1951}.
In Eq.\,\eqref{eq4}, the feedback coefficient, $\alpha/n$, is chosen
so as to satisfy (i) the summation about $n$ diverges, while (ii)
the sum of squares converges to a finite value.  Condition~(i) ensures that
$\theta^{(n)}$ can reach $\theta^*$ irrespective of the initial
value $\theta^{(0)}$, and condition~(ii) keeps the accumulated
variance to be finite.
Although the choice of $\alpha$ affects the convergence rate and the fluctuation of $\theta^{(n)}$ around $\theta^*$, the convergence is guaranteed as long as $\alpha$ is positive and finite.
Extension of this algorithm to higher dimensions is straightforward~\cite{AlbertG1970}.

\begin{figure}[t]
\begin{center}
\includegraphics[width=88mm]{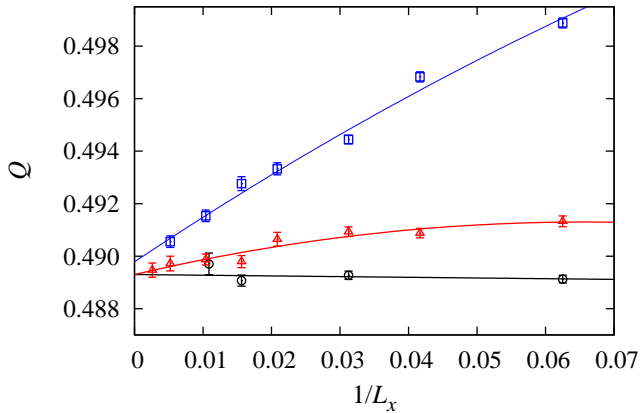}
\caption{System size dependence of the Binder ratio at the critical
  point for the the columnar (blue squares), staggered (red triangles)
  dimer models and the 3D classical Heisenberg model (black circles).  The coupling
  constant is fixed to the critical point, while the aspect ratio is
  optimized for each $L_x$ for the dimer models.}
\label{binder}
\end{center}
\end{figure}

At each Robbins-Monro step (RMS), we evaluate the correlation lengths
by 500 Monte Carlo steps with fixed parameters, $J^\prime$, $L_y$, and
$L_\tau$.  Then, the parameters are updated according to
Eq.\,\eqref{eq4}.  We choose the regression functions as
$R_x+R_y+R_\tau-3R_0$, $R_y-R_x$, and $R_\tau-R_x$, and the feedback
parameter $\alpha$ as 2, 500, and 500 for the parameters, $J^\prime$,
$L_y$, and $L_\tau$, respectively.  Here, it should be noted that the
quantum Monte Carlo simulation can be carried out only for integral
values of $L_y$, since $L_y$ is the number of lattice sites in
$y$-direction.  Furthermore, $L_y$ should be even in order to avoid negative signs.
For a non-even value of $L_y$ we adopt 
an arithmetic mean of two independent Monte
Carlo estimates for $L_y^{(1)}=W$ and $L_y^{(2)}=W+2$, where $W$ is
the maximum even integer not in excess of $L_y$.

Fig.\,\ref{params_multi} shows the convergence of the parameters for staggered dimer model with $L_x=64$.
One can see all $R_\alpha$'s oscillate coherently with decreasing amplitude from 20 to 50\,RMS, reflecting the convergence of $L_y$ and $L_\tau$.  In that region, $J^\prime$ still shows oscillatory behavior.
This means the gain for $J^\prime$ in the Robbins-Monro procedure is too large.
Although one might be able to optimize the gain to increase the convergence rate, such tuning only affects the number of steps before convergence.
For the present example (shown in Fig.\,\ref{params_multi}), the average is taken over 100 RMS with discarding first 300 RMS as the thermalization.  We have to discard more steps when the system size becomes larger. For example, we discard 650 RMS in the case of $L_x=384$.

After estimating $J^\prime_\rmc(L_x)$ for each $L_x$, then we extrapolate it in the thermodynamic limit $L_x\to\infty$.
We found $J^\prime_\mathrm{c}(L_x) = aL_x^{-b}+J^\prime_\mathrm{c}(\infty)$ is a good fitting function, where $a$, $b$, and $J^\prime_\mathrm{c}(\infty)$ are fitting parameters.
We used the data with the system size ranging from $L_x=16$ to $L_x=384$ for the staggered dimer model and to $L_x=192$ for the columnar dimer model. From this fitting, we conclude that the critical points $J^\prime_\mathrm{c}=2.51941(2)$ and $1.90947(3)$ for the staggered and columnar dimer model, respectively. They are consistent with those in the past literature, $J^\prime_\rmc=2.5196(2)$ for the staggered dimer model \cite{WenzelBJ2008} and 1.9096(2) for the columnar dimer model \cite{MatsumotoYTT2001, WenzelJ2009}, but our estimates are much more precise.

\begin{figure}[t]
\begin{center}
\includegraphics[width=88mm]{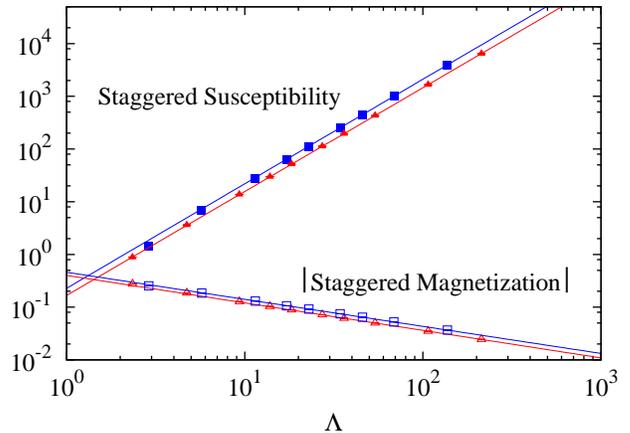}
\caption{System size dependence of the staggered susceptibility
  (filled symbols) and staggered magnetization (open symbols) at the
  critical point for the columnar (squares) and staggered (triangles)
  dimer models.  The coupling constant is fixed to the critical point,
  while the aspect ratio is optimized for each system size.  The lines
  are obtained by the least-squares fitting for the largest four
  system sizes.}
\label{exponents}
\end{center}
\end{figure} 

Let us move on to the calculation of the Binder ratio at the critical point.
For classical ferromagnetic Heisenberg models, the Binder ratio is defined as Eq.\eqref{eq.q},
where $m^z = \sum_{i} S_i^z$.
The quantum counterpart is defined in the same way except that $m_s^z$ is not the simple staggered magnetization (the N\'eel order parameter) but the integrated staggered magnetization, which is written as
\begin{equation}
m_s^z = \int_0^\beta \sum_{i} (-1)^{x_i+y_i}S^z(\tau)\diff\tau,
\end{equation}
as it reduces to $\sum_i S_i^z$ when one maps quantum systems into classical systems.
Fig.\,\ref{binder} shows the size dependence of the Binder ratio at the critical point calculated under the dynamic controlling of anisotropy.
Fitting the data with quadratic functions of $1/L_x$ gives the critical amplitudes in the thermodynamic limit as
\begin{equation}
Q_\rmc=
\begin{cases}
0.4893(2) \quad \text{for staggered dimer} \\
0.4898(3) \quad \text{for columnar dimer}\\
0.4893(2) \quad \text{for 3D classical.}
\end{cases}
\label{eq.5.7}
\end{equation}
We conclude that these three models share the same critical Binder ratio and thus they belong to the same 3D $O(3)$ universality class.
Note that these values are clearly different from that of other universalities, e.g., $Q_\rmc \simeq 0.62$ for the  3D Ising model.

We estimated the critical exponents as well.
The absolute value of the N\'eel order parameter $\left|m_s^z\right|$ and the staggered susceptibility $\chi$ behaves as $\langle \left|m_s^z\right| \rangle \sim \Lambda^{-\beta/\nu}$ and $\chi \sim \Lambda^{\gamma/\nu}$ at the critical point, respectively. Here, $\Lambda$ is the characteristic length of the system.
In the present simulation, $L_y$ and $L_\tau$ are optimally tuned so that the system should be virtually isotropic, thus we define $\Lambda$ as $\Lambda\equiv \left(L_x L_y L_\tau\right)^{1/3}$.
For the magnetization, fitting for the largest four data gives $\beta/\nu=0.522(3)$ and $0.513(9)$ for the staggered and columnar dimer model, respectively. Both coincide with the standard $O(3)$ value, 0.518(1) \cite{ChenF1993, CampostriniHPRV2002} and exclude the possibility of $\beta/\nu=0.545(4)$ from Ref.\,\citenum{WenzelBJ2008}.
For the staggered susceptibility, our estimate is $\gamma/\nu=1.970(5)$ and $1.983(13)$ for the staggered and columnar dimer model, respectively.  They are consistent with $\gamma/\nu= 1.9750(35)$ in Ref.\,\citenum{ChenF1993}, but is slightly bigger than $1.963(2)$ in Ref.\,\citenum{CampostriniHPRV2002}. In any case, we can see no evidence that the columnar and staggered dimer model belong to different universality classes.

\begin{figure}[tb]
\begin{center}
\includegraphics[bb=10 61 1008 751,width=85mm]{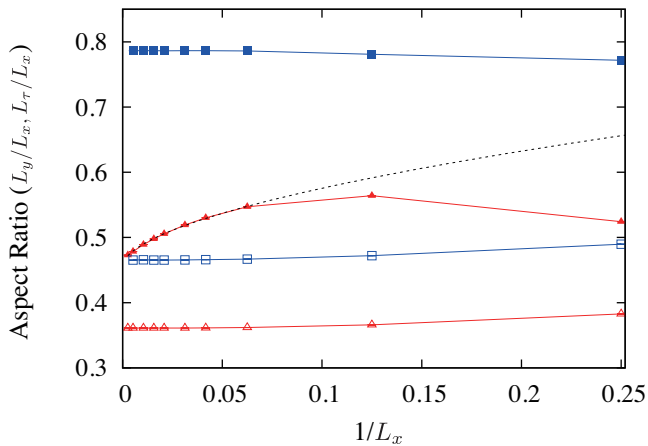}
\caption{System size dependence of the optimized aspect ratios, $L_y/L_x$ (filled symbols) and $L_\tau/L_x$ (open symbols), for the columnar (squares) and staggered (triangles) dimer models.
  The dashed line denotes the asymptotic behavior of $L_y/L_x$ for the staggered dimer model, $L_y/L_x \approx a + bL_x^{-\omega}$ with $\omega = 0.57$, obtained by the least-squares fitting for the largest four system sizes.}
\label{aspectratio}
\end{center}
\end{figure}

The apparent unconventional critical behavior observed in
Ref.\,\citenum{WenzelBJ2008} is attributed to the strong size
dependence of the virtual aspect ratio in the staggered dimer model.
Fig.\,\ref{aspectratio} shows the the optimized aspect ratio at the
critical point $J'=J'_\rmc$ as a function of the system size.  As one
can see, $L_y/L_x$ of the staggered dimer model exhibits a strong and
non-monotonic size dependence, though the other ratios converge to
finite values quite rapidly.  This fact means that in the conventional
simulation with a fixed aspect ratio $L_x : L_y : L_\tau$, the virtual
aspect ratio $R_x^{-1} : R_y^{-1} : R_\tau^{-1}$ gradually changes as
the system size increases that introduces strong corrections to
scaling in the staggered dimer model.

Let us analyze this unconventional behavior in a different way.
From the quantum field theory, the low-energy effective action for the present models is described by the standard $\phi^4$ action $\mathcal{S}_0 \sim (c_x\partial_x\vec{\phi})^2+(c_y\partial_y\vec{\phi})^2+(c_\tau\partial_\tau\vec{\phi})^2+m\vec{\phi}^4$, where the constants which determine the length scale are left explicitly as $c_\alpha$.
Along this action, only for the staggered dimer model, it is discussed that the action has the cubic term $\mathcal{S}_3\sim\gamma\vec{\phi}\cdot(\partial_x\vec{\phi}\times\partial_\tau\vec{\phi})$ \cite{FritzDWWBV2011}.
It is straightforward to show that rotating in the $x$-$\tau$ plane effectively pushes $\mathcal{S}_3$ into the kinetic term of the action $\mathcal{S}_0$, resulting that the coefficients are renormalized as $c_x, c_\tau \sim 1+L_x^{-[\gamma\phi]}$, where $[\,\,\cdot\,\,]$ denotes the scaling dimension of $\cdot$, whereas $c_y$ remains unchanged.
This simple dimensional analysis explains the behavior of the optimized aspect ratio of the staggered dimer model observed in Fig.\,\ref{aspectratio}, where $c_y/c_x$ suffers from large corrections but $c_\tau/c_x$ does not.
We assume the form of finite size correction as $L_y/L_x \approx a+bL_x^{-\omega}$.  The exponent of correction, $\omega = 0.57(4)$, obtained by least-squares fitting is consistent with the above estimate assuming that 
$\mathcal{S}_3$ is weakly irrelevant, i.e., $[\gamma]$ is negative and has small absolute value, and $[\phi]\approx 0.5$.  Our preliminary simulation for the herringbone dimer model (see Fig.\,1(c) in Ref.\,\citenum{FritzDWWBV2011}) that is considered to have the cubic term shows a similar unconventional behavior of the aspect ratio as the staggered dimer model~\cite{YasudaT2100}.

In this paper, we presented the finite size scaling method with controlling anisotropy of the system dynamically. In virtually isotropic systems, the corrections to scaling peculiar to the anisotropic systems is reduced, and we can compare the critical amplitudes among the classical and quantum systems.
This method can give the optimal system size including $L_\tau$, which has been chosen sufficiently large value because there has been no index.
We applied this method to the spatially anisotropic Heisenberg models, which was considered to be hard to judge its universality class because of the extremely large corrections to scaling. We concluded they belong to the same standard $O(3)$ universality class based on the critical amplitudes and the critical exponents. We also revealed the optimized aspect ratio shows the non-monotonic behavior from the cubic term $\mathcal{S}_3$ of the effective action.

The simulation code has been developed based on the ALPS/looper library\cite{ALPS2011s, ALPSweb,TodoK2001}. I acknowledge support by the Grand Challenge to Next-Generation Integrated Nanoscience, Development and Application of Advanced High-Performance Supercomputer Project from MEXT, Japan, the HPCI Strategic Programs for Innovative Research (SPIRE) from MEXT, Japan, and the Computational Materials Science Initiative (CMSI). 

\bibliography{main}

\end{document}